# Chemical Functionalization of Graphene Nanoribbons by Carboxyl Groups on Stone-Wales Defects

Fangping OuYang,[1] Bing Huang,[2] Zuanyi Li,[2] and Hui Xu[1,†]

[1]School of Physics Science and Technology, Central South University, Changsha 410083, P. R. China

[2]Department of Physics, Tsinghua University, Beijing 100084, P. R. China

(oyfp04@mails.tsinghua.edu.cn)

Using the density functional theory, we have demonstrated the chemical functionalization of semiconducting graphene nanoribbons (GNRs) with Stone-Wales (SW) defects by carboxyl (COOH) groups. It is found that the geometrical structures and electronic properties of the GNRs changed significantly, and the electrical conductivity of the system could be considerably enhanced by mono-adsorption and double-adsorption of COOH, which sensitively depends upon the axial concentration of SW defects COOH pairs (SWDCPs). With the increase of the axial concentration of SWDCPs, the system would transform from semiconducting behavior to $p$-type metallic behavior. This fact makes GNRs a possible candidate for chemical sensors and nanoelectronic devices based on graphene nanoribbons.

# I. Introduction

Chemical functionalization of low-dimensional carbon network systems such as carbon nanotubes（CNTs）, graphene and graphene nanoribbons (GNRs), by introducing molecules and groups has attracted a lot of attention as a significant way to modify their physical and chemical properties[1-4]. Stable defects in a graphene layer, such as topological defects[5.6], vacancies[7–13], and adatoms[14-17], have been experimentally and theoretically proved to be numerous. The Stone-Wales (SW) defect is a typical topological defect in the carbon nanostructures and is comprised of two pairs of five-membered and seven-membered rings. It was suggested that SW defects could act as nucleation centers for the formation of dislocations in the originally ideal graphite network and constitute the onset of the possible plastic deformation of a CNT[18]. Ayako Hashimoto et al. have reported the observations in situ of defect formation in single graphene layers by high-resolution TEM [19]. We can envisage more diversified applications in carbon nanomaterials by taking advantage of these defects, which can be induced locally during electron irradiation.

Recently, chemical adsorption such as $H_2$、OH、$H_2O$、CO$x$ and NO$x$ on pristine and defective graphene have attracted special attention both experimentally and theoretically [3,4,6, 20-22]. As a main prototype for chemical functionalization, the carboxyl groups (COOH) were usually found on the carbon materials after the oxidation of carbon by strong acids[23-24], or other oxidizing agents[25-26]. In addition, such oxidized carbon materials can be provided with a variety of functional moieties for further functionalization, since the derivative COOH groups are able to couple molecules with amide and ester bonds[27-29]. Thus, it is more interesting and practical to obtain more comprehensive information on the physical and chemical properties of defective GNRs with COOH groups and to predict possible applications.

In this letter, using first-principles calculations, we study the chemical functionalization on GNRs with SW defects by COOH groups. Our main results are the following. (i) We explore the nature of the interaction between COOH groups and SW defects and explain the associated adsorption behavior from the localized electronic states. (ii) We evaluate the effect of functionalization of GNRs by introducing COOH groups on SW defects and predict possible applications in chemical sensors and nanobioelectronics.

## II. Calculation Method and Model

There are two kinds of typical GNRs -- armchair edges ribbons (AGNRs) and zigzag edges ribbons (ZGNRs). All ZGNRs are metallic at finite temperature, and AGNRs are either metallic or semiconducting depending on the ribbon direction and width that is the same as those of CNTs on the tube diameter and chirality[30-35]. We choose armchair (8,3) nanoribbons as the typical supercell (as shown in Figure 1). Herein, parameter 8 is defined as the number of carbon atom along the ribbon width and parameter 3 represents the number of prime cells.

The structural optimizations and electronic structure calculations were performed using the Vienna Ab initio Simulation Package (VASP)[36] in the formalism of planewave ultrasoft pseudopotential [37] within the local density approximation （LDA）[38]. The Ceperley-Alder exchange-correlation energy functional was used. The plane-wave cutoff energy is 350 eV for all calculations. All of the structures were fully optimized using the conjugate gradient algorithm until the residual forces was smaller than 10 eV/Å. The edges of GNRs are all H-terminated so as to remove dangling bonds. To explicitly study the interaction between COOH groups and a SW-defective (8,3) GNR(as shown in Figure 1a), we employ a supercell of 10 Å x10 Å x12.681 Å, containing 96 C atoms and one COOH group (mono-adsorption mode as shown in Figure 1b) or two COOH groups (double adsorption mode), together with periodic boundary conditions. The Brillouin-zone integration is performed with 1 x 1 x 11 Monkhorst-Pack **k** points. All parameters for calculation are choosed after testing and ensure a good enough convergence of total energies.

## III. Results and Discussion

For the low-dimensional carbon *sp*2 network systems such as CNTs and GNRs, it is found that the donor states induced by the SW defect are advantageous for the adsorption of the COOH groups or oxidation. Especially, in common, the two atomic sites at the center of the SW defect are the most energetically favorable adsorption sites.[39] (C1, C1′atoms as shown in Figure 1a).

Firstly, we investigated the structural properties of the system in which the COOH group is attached to the C1 site of the SW defect. As shown in Fig. 1b, the COOH group pulls out the C1 atom a little from the ribbon surface, leading to a significant change of local geometry of the SW defect. The forming C-COOH bond is nearly perpendicular to the ribbon surface, and the three associated angles are 109.5, 109.5, and 105.6°, respectively, as shown in Fig.2c. For the SW defect on the ribbon, the C-C bond lengths are increased from 1.46, 1.46, and 1.32 Å (shown in Figure 2a) to 1.53, 1.53, and 1.44 Å (shown in Figure 2b), and correspondingly, the bond angles are reduced from 114.7, 122.7°, and122.7° (shown in Figure 2a) to 108.1, 110.6°, and 110.8° (shown in Figure 2b). These C-C bond angles significantly deviate from the standard angle of sp2 hybridization, 120° which indicates that the carbon atoms of the SW defect near the adsorption site, C1, change to sp3 hybridization from sp2 hybridization (shown in Figure 1c). With the near-sp3 hybridized bond between the COOH group and the ribbon, a quasitetrahedral bonding configuration of carbon atoms such as the methane molecule, CH4, is formed at the adsorption site. Moreover, the binding energy ($E_b$) of the COOH group attached to the ribbon is found to be -2.78 eV which indicate that the adsorption process is chemical in nature.. Herein, the binding energy $E_b$, is defined as $E_b = E_t(GNR+COOH) - E_t(GNR) - E_t(COOH)$, where $E_t(GNR + COOH)$, $E_t(GNR)$, and $E_t(COOH)$ are the total energies of the tube with the COOH group, of the ribbon, and of the COOH group, respectively.

In the following, we study in detail the electronic properties of this system in which the COOH group is attached to the C1 site of the SW defect. Figure 3 shows the band structures and the local density of states of defective armchair (8, x) (x=2, 3, 4) GNRs before and after COOH group adsorption with several deferent axial densities of SW defects COOH pairs (SWDCPs). Our calculation also shows that in the semiconducting ribbon, the conductivity could be considerably enhanced via the formation of SWDCPs. In Fig. 3(a), we can see that for the pristine SW-GNR, a defect band（flat band） is present at about 0.6 eV above the Fermi level, and from partial density of states (PDOS) analysis, we found that this defect band is mainly contributed by C1 atom. Then after the COOH adsorption, the above defect band substantially shifts down to the Fermi level, as shown in Figs. 3(b)–3(d). such a shift down of the band becomes more considerable along with the increase of the axial concentration of SWDCPs, and finally the system is transformed from semiconducting behavior to p-type metallic behavior. For example, in the case with the SWDCP density of 0.059 Å−1, the half-filled dispersionless band is very close to the highest valence band Fig. 3(d)—which corresponds to a shallow localized impurity state and indicates a conventional *p*-type semiconducting behavior. Thus, the whole system may have much better conducting behavior, which is substantially favorable to the electrical conductivity of COOH-functionalized SW-GNR. In addition, with the increase of the axial concentration of SWDCPs, the band gap the system is reduced from 0.67 eV of the pristine SW-defective GNR (Fig. 3a) to 0.50 eV of the SW-defective one with adsorption of the COOH group (Fig. 3d). It shows delocalized electronic states over the ribbon, provides an effective conducting channel, which is substantially favorable to the electrical conductivity of COOH-functionalized SW-GNRs.

Next, we study the structural and electronic properties of the system in which the COOH groups is double adsorbed on the C1 and C1′ sites of the SW defect. As shown in Fig.4, the spatial orientations of the two adsorbed COOH groups are almost "antiparallel" with the angle between the tube axis and COOH groups being about 30.6° The C1 and C1′ atoms in the SW defect are pulled a little out of the ribbon plan, and the C1-C1′ bond length increases to 1.47 Å from 1.32 Å in the pure defective tube. The formed C-COOH bonds with the identical length of 1.55 Å remain nearly perpendicular to the tube axis, but the three associated angles change to 110.6, 109.1, and 95.17°, respectively, as shown in Figure 4d. The two individual COOH groups have the same deformation effect on the SW defect, and the resulting geometry of the deformed SW defect (Fig. 4b) is the following: the C-C bond lengths are 1.47, 1.52, and 1.54 Å, and the bond angles are116.5, 116.2, and 107.3°. This indicates that similar to the case of mono-adsorption, the quasitetrahedral bonding configurations of carbon atoms based on sp3 hybridization are formed at the C1 and C1′ sites in the double-adsorption defective GNRs.

Figure 5 shows band structures of the perfect、pristine、 SW-defective、mono-adsorption and double-adsorption SW-defective GNRs. As shown in Figure 5b （or Figure 3a）, the SW defect induces a significant change of the electronic properties of the GNR. There is a defect states (marked by the black arrow) introduced on the bottom of conduct band level. Similar to the case of mono-adsorption, the double adsorption of COOH groups significantly changes electronic properties of the defective GNRs, as shown in Fig. 5c and Fig. 5d.

With the energy band theory, the chemical functionalization mechanism for such system can be interpreted. The unsaturated electron of the carbon atom in the COOH group directly interacts with the donor states localized at the C1 site, when the COOH group is attached to the SW defect. The strong rehybridization interaction causes obvious downward shifts of the states from the two bonding C atoms to the lower energy region. This can explain the phenomenon that the defect band in Figure 5b shifts down after the adsorption of the COOH group (in Figure 5c-d). Essentially, the redistribution of the electronic states is responsible for the change of electrical conductivity with the density of SWDCPs. In the case of higher density, SWDCPs can interact with each other through the redistributed electronic states, leading to stronger bonding between the COOH group and SW atoms.

We can expect that as the COOH groups are readily derivatized by a variety of reactions, the preparation of a wide range of functionalized GNRs with topological defects should be possible, which may make GNRs equipped with a variety of functional moieties as potentially useful building blocks for nanoelectronic and nanobioelectronic devices. Also, this finding would provide an approach, chemical oxidation, to tailor the electronic properties of semiconductor GNRs.

## IV. Conclusion

In summary, we have systematically investigated the mechanism and effect of COOH functionalization of SW-GNRs, using the density functional theory. The interactions between SW defects and COOH groups would cause the changes of geometries and electronic structures of SW-GNRs. The COOH group leads to the significant change of local geometry of the SW defect, and a quasitetrahedral bonding configuration of carbon atoms is formed in the adsorption site, indicating sp3 hybrid bonding. On the other hand, our calculation shows the electrical conductivity of the system would be significantly improved by increasing the axial concentration of SWDCPs for mono-adsorption and double adsorption of COOH on the defective GNRs. With the increase of the axial concentration of SWDCPs, the system transformed from semiconducting behavior to p-type metallic behavior finally. Our work suggests semiconducting GNRs could be a possible candidate for chemical sensors and nanoelectronic devices under some special conditions, such as in aqueous or organic solvents, since COOH groups can bind with other molecules through various reactions. Thus, GNRs equipped with a variety of functional moieties could be potentially useful as building blocks for nanoelectronic and nanobioelectronic devices. We hope that our observations may stimulate further experimental analysis.

**Acknowledgment.**

This work was supported by the National Natural Science Foundation of China (Grant Nos. 10325415 and 50504017). The numerical calculation was carried out by the computer facilities at Department of Physics of Tsinghua University.

**Figure Captions**

**FIG 1.** (Color online) (a) Optimized morphologies of SW defect (in black) in the armchair (8,3) graphene nanoribbons. (b) Optimized morphologies of the SW defect defective armchair (8,3) nanoribbon with the COOH group. (c) Atomic structure of the SW defects with the COOH group. The grey, red, and white balls correspond to carbon, oxygen, and hydrogen atoms, respectively.

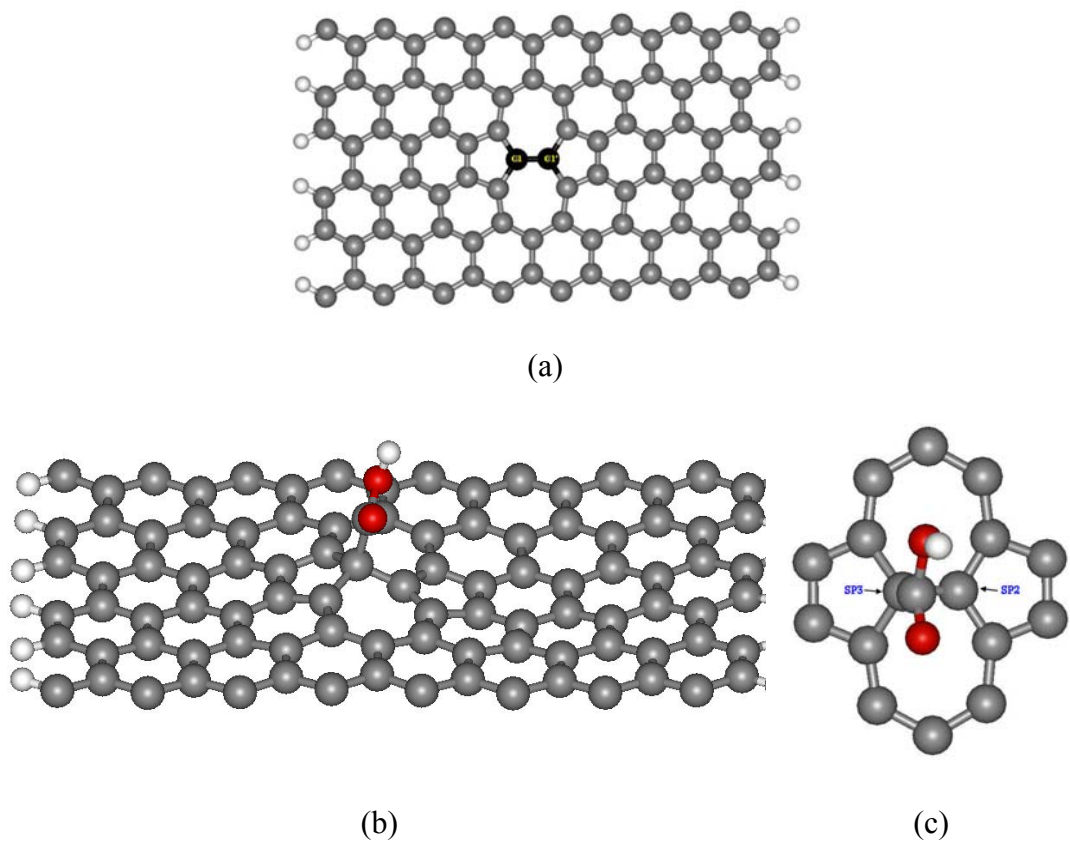

(a)

(b) (c)

**FIG 2.** Optimized structures of (a) the SW defect and (b) the SW defect with the COOH group on the armchair (8,3) GNR. (c)The associated bond lengths and bond angles between the COOH group and the SW defect. The black ball represents associated carbon atom.

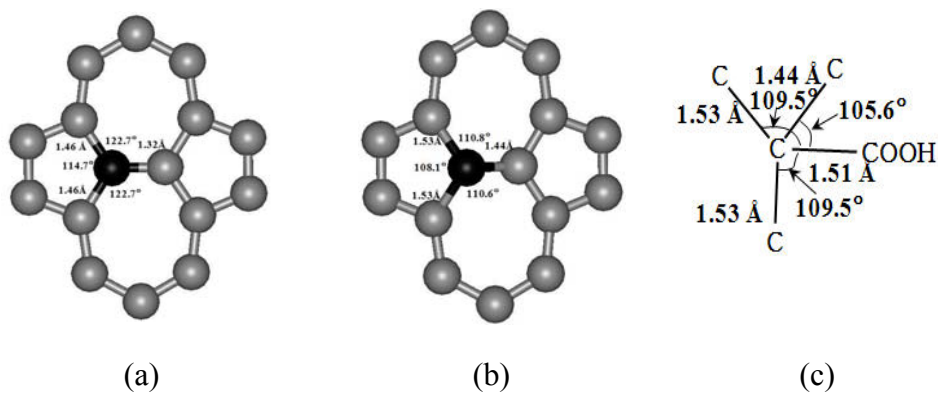

(a)  (b)  (c)

**FIG 3.** (Color online) Band structures and density of states (DOS) of (a) pristine SW-defective armchair (8,3) GNR without adsorption of the COOH group, with the axial concentration of SW-defects of 0.079 Å$^{-1}$, corresponding to the super-cell size of 12.681 Å along the $z$ axis. (b)–(d) COOH-functionalized GNRs with various axial concentrations of SW-defects: (c) 0.079 Å$^{-1}$, (d) 0.059 Å$^{-1}$, and (e) 0.118 Å$^{-1}$. The Fermi level is indicated with red dashed line. "Total," "C1," and "COOH" represent the DOS of the whole system, the carbon atom at C1 site, and the COOH group, respectively.

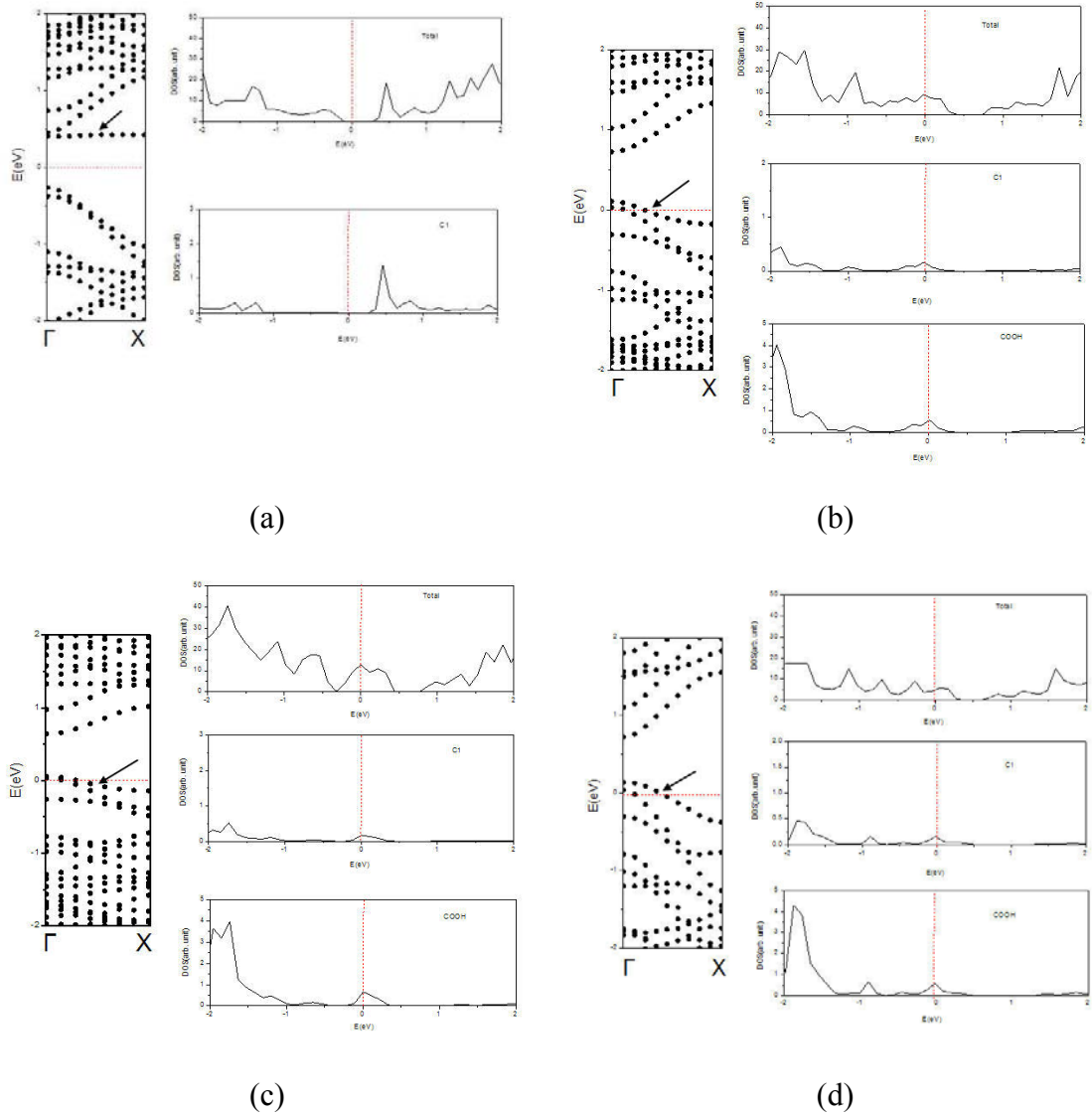

(a)  (b)

(c)  (d)

**FIG 4.** (Color online) (a)-(c) Atomic structure of the double-adsorption defective GNRs. The grey, red, and white balls correspond to carbon, oxygen, and hydrogen atoms, respectively. (d) Optimized morphologies of the SW defect with two COOH groups. (e) The associated lengths and angles of the C-COOH bond in the double adsorption.

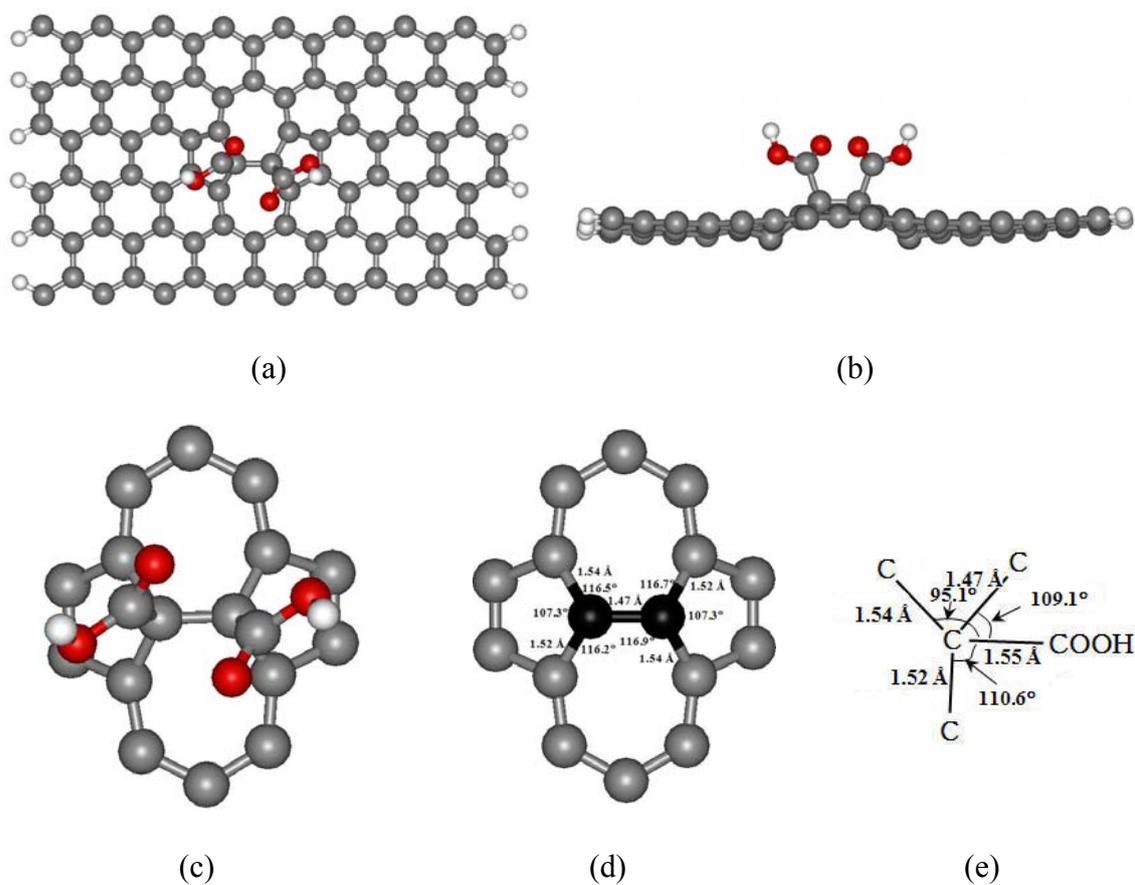

(a)　　　　　　　　　　(b)

(c)　　　　　　(d)　　　　　　(e)

**FIG 5**. Band structures of the (a) perfect (b) Pristine SW-defective (c) mono-adsorption and (d) double-adsorption SW-defective GNRs. The Fermi level is indicated with red dashed line.

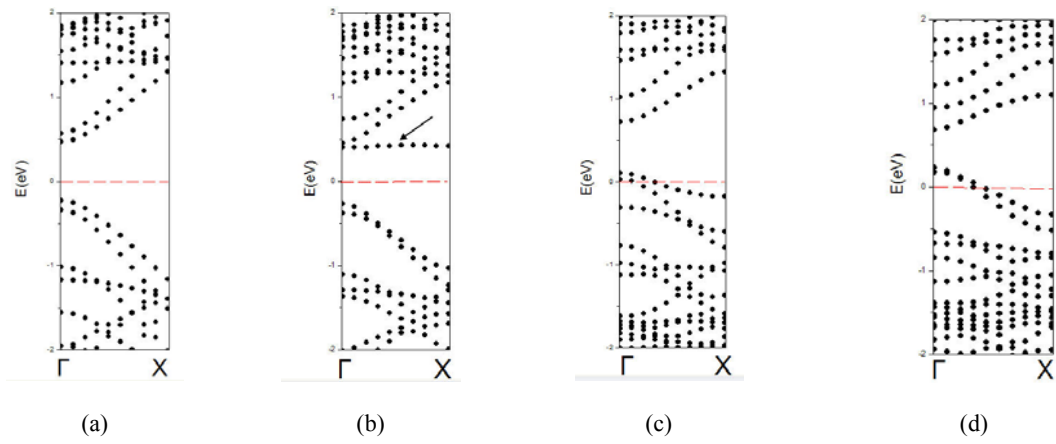

(a)　　　　　(b)　　　　　(c)　　　　　(d)